\documentclass[12pt, a4paper]{article}
\pdfoutput=1
\usepackage{a4wide}
\usepackage{amsfonts}
\usepackage{amscd}
\usepackage{amssymb}
\usepackage{amsmath}
\usepackage{epsfig}
\usepackage{latexsym}
\usepackage{stmaryrd}
\usepackage{hyperref}
\usepackage{setspace}
\hypersetup{
    colorlinks=true,
    linkcolor=black,
    citecolor=black,
    filecolor=black,
    urlcolor=black,
}

\def \be  {\begin{equation}}
\def \ee  {\end{equation}}
\def \ba  {\begin{eqnarray}}
\def \ea  {\end{eqnarray}}
\def \bb  {\begin{thebibliography}}
\def \eb  {\end{thebibliography}}
\def \lab #1 {\label{#1}}

\newcommand\tr{\mathrm{Tr}}

\newcommand{\captionfonts}{\small}

\makeatletter 
\long\def\@makecaption#1#2{%
\vskip\abovecaptionskip
\sbox\@tempboxa{{\captionfonts #1: #2}}%
\ifdim \wd\@tempboxa >\hsize
  {\captionfonts #1: #2\par}
 \else
   \hbox to\hsize{\hfil\box\@tempboxa\hfil}%
  \fi
  \vskip\belowcaptionskip}
\makeatother

\begin{document}

\thispagestyle{empty}
\vskip1.5truecm
\begin{center}
\vskip 0.2truecm 

{\Large\bf On S-duality of the Superconformal Index on Lens}
\vskip 3truemm
{\Large\bf Spaces and 2d TQFT}
\vskip 1truecm
{\bf Luis F. Alday, Mathew Bullimore and Martin Fluder\\}

\vskip 1cm

\textit{Mathematical Institute, University of Oxford,\\
24-29 St Giles', Oxford OX1 3LB, United Kingdom\\}

\vskip 0.8cm

\end{center}

\vskip 2 cm

\begin{abstract}
\noindent We consider the 4d superconformal index for ${\cal N}=2$ gauge theories on $S^1 \times L(r,1)$, where $L(r,1)$ is a Lens space. We focus on a one-parameter slice of the three-dimensional fugacity space and in that sector we show S-duality. We do so by rewriting the index in a way that resembles a correlation function of a 2d TFT, which however, we do not identify. 

 \end{abstract}

\newpage

\section{Introduction}

Over the last few years we have seen a huge progress in understanding ${\cal N}=2$ super-symmetric theories in four dimensions. Much of this progress was due to the realization that we could systematically construct such four dimensional theories by compactifying the six-dimensional theory living of $N$ M$5-$branes (the putative ${\cal N}=(2,0)$ six-dimensional theory of type $A_{N-1}$) on a Riemann surface with punctures \cite{Gaiotto:2009we}. Furthermore, the properties of the resulting 4d theory are in correspondence with the properties of the Riemann surface. For instance, S-duality corresponds to different pant decompositions of the same Riemann surface. This suggests that the physical quantities of the 4d theory should be closely related to those of the theory on the two-dimensional Riemann surface. The first example of this relation is the correspondence between the partition function on $S^4$ and correlators in 2d Liouville  CFT \cite{Alday:2009aq}.

Another powerful relation, more relevant to this paper, involves the super-conformal index, or the super-symmetric partition function on $S^1 \times S^3$. It was found in \cite{Gadde:2009kb} that invariance of the index under S-duality translates into associativity for the operator algebra of a 2d TQFT. Furthermore, for a particular choice of the fugacities, this 2d theory was identified as $q-$deformed 2d Yang-Mills in the zero area limit \cite{Gadde:2011ik}. Extensions to other choices of fugacities were studied in \cite{Gadde:2011uv}.

On general grounds, the relation between the super-conformal index and $q-$deformed Yang-Mills can be understood as follows. We think of the partition function on $S^1 \times S^3$ as arising from the compatification over $\Sigma$  of an observable on the six dimensional ${\cal N}=(2,0)$ theory on $S^1 \times S^3 \times \Sigma$. On the other hand, we can also compactify the six-dimensional theory on $S^1 \times S^3$, leading to an observable in a two dimensional theory on the Riemann surface $\Sigma$. If the 6d observable is ``protected", this leads to a relation between the observables in the 4d and 2d theories. By analogous reasoning, considering the 6d theory on $S^4 \times \Sigma$, leads to the relation mentioned above, between the partition function on $S^4$ and Liouville correlators. 

The above idea is hard to make precise, since little is known about the six-dimensional theory. However, in the case of the super-conformal index, the presence of the $S^1$ makes things simpler: the six dimensional theory compactified over $S^1$ gives rise to five-dimensional ${\cal N}=2$ Yang-Mills theory \cite{Douglas:2010iu,Lambert:2010iw}, so we could try to understand the super-conformal index/2d TQFT duality by studying the 5d theory on $S^3 \times \Sigma$. Indeed, in \cite{Fukuda:2012jr} the partition function of the 5d theory on $S^3 \times \Sigma$ was computed and it was found to be identical to the partition function of the 2d q-deformed Yang-Mills theory. It is tantalizing then to conjecture that in general, the super-symmetric partition function of a 4d ${\cal N}=2$ theory on $S^1 \times M_3$ will be given in terms of some 2d TQFT, given by the compactification of  five-dimensional ${\cal N}=2$ Yang-Mills theory on $M_3$. 

In this paper we make a step towards this direction by studying the super-symmetric partition function on $S^1 \times L(r,1)= S^1 \times S^3/\mathbb{Z}_r$. This index was computed in \cite{Benini:2011nc}, where $S-$duality was checked up to a few orders in a fugacities expansion.  The aim of this paper is to prove S-duality (in a subspace of the fugacities space) by following an strategy similar to \cite{Gadde:2011ik}: We uncover the structure behind S-duality by rewriting the index in a way that resembles the correlation function of a 2d TFT. The structure of the index on Lens spaces however, is much more complicated that the structure of the index for $r=1$, and we are not able to identify the relevant 2d TFT.

The organization of this paper is as follows. In section two we review the computation of the index on $S^1 \times L(r,1)$. In section three we define the particular slice to be studied and we show that for this slice the index can be written in a way that resembles a correlator of a 2d TFT. Furthermore, we show that correlators of the putative 2d TFT satisfy crossing symmetry, which translates into S-duality for the index computation. In section four we end with some conclusions and open problems.

\section{The Index on $S^1 \times L(r,1)$}

The index is defined as \cite{Kinney:2005ej}

\begin{equation}
{\cal I} = \tr(-1)^F p^{j_1+j_2-r} q^{-j_1+j_2-r} t^{R+r}
\end{equation}
where $F$ is the fermion number and the trace is taken over the states of the theory on $S^3$ satisfying $E-2j_2-2R+r=0$. $E$ stands for the conformal dimension, $(j_1,j_2)$ for the Cartan generators of the $SU(2)_1 \otimes SU(2)_2$ isometry group and $(R,r)$ for the Cartan generators of the $SU(2)_R \otimes U(1)_r$ $R-$symmetry. 

The index on  $S^1 \times L(r,1) $ was computed in \cite{Benini:2011nc}, to where we refer the reader for the details. The Lens space $L(r,1)=S^3/\mathbb{Z}_r$ is defined as the orbifold of $S^3: \{ (z_1,z_2) ~\epsilon~ \mathbb{C}^2, |z_1|^2+|z_2|^2=1 \}$ under the identification 

$$(z_1,z_2) \sim (e^{2\pi i/r} z_1,e^{-2\pi i/r} z_2)$$
where $SU(2)_1$ acts on $(z_1,z_2)$ as a doublet. $\mathbb{Z}_r$ acts on the $S^1$ fiber of the Hopf fibration, denoted by $S^1_H$, {{\it} i.e.} the $\mathbb{Z}_r$ action is embedded into $U(1)_i \subset SU(2)_1$.

The orbifold theory has a set of vacua labeled by a non-trivial holonomy $V$ along the $S^1_H$, with $V^r=1$. In this paper we will restrict to the case of $A_1$-type theories in which case the holonomy can be taken of the form

\begin{equation}
V=diag(e^{2\pi i m/r},e^{2\pi i (r-m)/r}),~~~m=0,1,...,[r/2].
\end{equation}
Different sectors are then labeled by the integer $m$. 

The total index of a theory is constructed from the following building blocks, which could be interpreted as three-point functions and propagators:

\begin{eqnarray*}
{\cal I}_{trif}^{(m_1,m_2,m_3)}(a_1,a_2,a_3) = {\cal I}^{(m_1,m_2,m_3)}_0 \exp \left({\sum_{{\bf s}}\sum_{n=1}\frac{1}{n} g(t^n,p^n,q^n,[{\bf m} . {\bf s} ]_r) a_1^{n s_1} a_2^{n s_2} a_3^{n s_3}} \right) \\
\eta^{(m)}(a)=\eta^{(m)}_0 \exp \left( {\sum_{n=1} \frac{1}{n} \left( f(t^n,p^n,q^n;[2m]_r)a^{2n}+ f(t^n,p^n,q^n,[-2m]_r)a^{-2n}+ f(t^n,p^n,q^n;0) \right)} \right)
\end{eqnarray*}
where $[x]_r$ denotes $x~mod~r$ and we have introduced ${\bf s}=(s_1,s_2,s_3)$, $s_i=\pm$ and ${\bf m}=(m_1,m_2,m_3)$. The zero-point contributions are given by 

\begin{equation}
{\cal I}^{(m_1,m_2,m_3)}_0 = \left(\frac{\sqrt{t}}{p q} \right)^{-\frac{1}{4} \sum_{{\bf s}}([{\bf m} . {\bf s} ]_r-[{\bf m} . {\bf s} ]_r^2/r)},~~~~~~\eta^{(m)}_0 = \left(\frac{\sqrt{t}}{p q} \right)^{[2m]_r-[2m]_r^2/r}
\end{equation}
Finally, the one letter contributions are given by
\begin{eqnarray}
f(t,p,q;m)&=& \frac{\frac{p^m}{1-p^r}+\frac{q^{r-m}}{1-q^r}}{1-p q}\left(p q(1+1/\sqrt{t}) -1 -\sqrt{t} \right)+\delta_{m,0}\\
g(t,p,q;m)&=&\frac{\frac{p^m}{1-p^r}+\frac{q^{r-m}}{1-q^r}}{1-p q} \left( t^{1/4} - \frac{p q}{t^{1/4}}\right)
\end{eqnarray}
The total index is then obtained by "gluing" such contributions. For instance, the "four-point function" is obtained by joining two three-point functions with the corresponding propagators

\begin{equation}
\label{four}
\sum_{m=0}^{[r/2]} \int [dz]^m {\cal I}_{trif}^{(m_1,m_2,m)}(a_1,a_2,a)  \eta^{(m)}(a) {\cal I}_{trif}^{(m,m_3,m_4)}(a,a_3,a_4) 
\end{equation}
where the measure is given by

\begin{equation}
[da]^m=\left\{
\begin{array}{c l}      
    \frac{2-a^2-a^{-2}}{4\pi} \frac{da}{a} & [2m]_r=0\\
    \frac{1}{4\pi} \frac{da}{a} & [2m]_r \neq 0
\end{array}\right.
\end{equation}

$S-$duality implies that the four-point function (\ref{four}) is symmetric under the interchange of any two pairs $(a_i,m_i) \leftrightarrow (a_j,m_j)$. This was verified to a few orders in the fugacities expansion in  \cite{Benini:2011nc}. In order for this crossing property to work, one needs to sum over all intermediate values of $m$. The aim of this paper is to understand such a property as arising from a 2d TQFT computation, as it was done for the $r=1$ case in \cite{Gadde:2009kb} and subsequent papers.

\section{A particular slice and 2d TFT picture}

\subsection{A particular slice}

 In \cite{Gadde:2011uv} a particular limit, denoted as the "Macdonald Index", was studied. This limit corresponds to  $p=0$, general $q,t$ and is characterized by its enhanced supersymmetry. Since only states with $j_1+j_2-r=0$ will contribute to the index in this limit, we expect a simplification. This can also be seen as follows. The building blocks of the index can be written as products of elliptic gamma functions of the form $\Gamma(z,p,q)$, where $z$ is some expression depending on the $a_i$ and the fugacities. The following identity
 
\begin{equation}
\Gamma(z,0,q)=\frac{1}{(z;q)},~~~~(z;q)=\prod_{i=0}^\infty(1-z q^i)
\end{equation}
Then implies that the index can be written in terms of simpler functions. Indeed, in this limit we find the following expressions

\begin{eqnarray}
\label{Cvspoch}
{\cal I}_{trif}^{(m_1,m_2,m_3)}(a_1,a_2,a_3) =  {\cal I}^{(m_1,m_2,m_3)}_0 \prod_{\bf s} \frac{1}{(\sqrt{t} a_1^{s_1}a_2^{s_2}a_3^{s_3} q^{[-m.s]_r};q^r)}\\
\eta^{(m)}(a)=\left(\frac{-a^2}{(1-a^2)^2} \right)^{\delta_{m,0}} (q^r;q^r)(t;q^r)\prod_{s=\pm} (a^{2s} q^{[-2m s]_r};q^r) (a^{2s} t q^{[-2m s]_r};q^r)
\end{eqnarray}
The zero-point contributions will be discussed momentarily. Mimicking  \cite{Gadde:2009kb}, we define the rescaled structure constants
\begin{equation}
\hat C^{(m_1,m_2,m_3)}(a_1,a_2,a_3) = \sqrt{\eta^{(m_1)}(a_1)\eta^{(m_2)}(a_2)\eta^{(m_3)}(a_3)} {\cal I}_{trif}^{(m_1,m_2,m_3)}(a_1,a_2,a_3)
\end{equation}
Note that the zero-point contributions are subtle in this limit, since they are proportional to $p$ to some power. It can be checked that this power is always bigger than or equal to zero. Furthermore, quite remarkably, this power is exactly zero (and so the contribution does not vanish in the $p \rightarrow 0$ limit), provided $m_1,m_2$ and $m_3$ satisfy a selection rule. Namely, for fixed $m_1$ and $m_2$, $m_3$ should run between $|m_1-m_2|$ and $min(|r-m_1-m_2|,m_1+m_2)$. Note that this agrees with the selection rules for $SU(2)$ affine algebra at level $r$.

In what follows, we will make a further simplification in the space of fugacities, and we will consider the limit $t=q^r$. Note that this reduces to the 'Schur' limit for $r=1$ and so the 2d-TFT we are after should reduce to $q-$deformed 2d YM in the zero area limit \cite{Gadde:2011ik}. 

\subsection{$2d$ TFT interpretation}

We would like to interpret the rescaled structure constants as the three-point correlation functions of some $2d$- TFT. Let us start by defining 

\begin{equation}
C^{(m_1,m_2,m_3)}(a_1,a_2,a_3)  \equiv \frac{1-q^r}{(q^r;q^r)}  \hat C^{(m_1,m_2,m_3)}(a_1,a_2,a_3) 
\end{equation}
and let us study this object for cases of increasing difficulty. The simplest case corresponds to $m_1=m_2=m_3=0$. One can explicitly check that in this case the structure constants coincide with that of the $r=1$ case, up to a rescaling $q \rightarrow q^r$. Using the results of \cite{Gadde:2011ik} we can immediately write

\begin{equation}
\label{qdef}
C^{(0,0,0)}(a_1,a_2,a_3)  = \sum^{\infty}_{\ell=1}\frac{\chi_\ell(a_1)\chi_\ell(a_2)\chi_\ell(a_3)}{|\ell|_{q^r}}
\end{equation}
where we introduced the Schur polynomials and the $q-$deformed dimension

\begin{equation}
\chi_\ell(a) = \frac{a^\ell-a^{-\ell}}{a-1/a},~~~~~~|\ell|_{q} = \frac{q^{-\ell/2}-q^{\ell/2}}{q^{-1/2}-q^{1/2}}
\end{equation}
The next case, allowed by the selection rules, corresponds to $m_1=0$ and $m_2=m_3=m$, with $0<m<r/2$. Note that in this case the three-point function is not symmetric under the interchange $a_2 \leftrightarrow 1/a_2$ or $a_3 \leftrightarrow 1/a_3$. Hence, we are not able to expand it purely in terms of Schur polynomials. However, we can expand it in terms of Schur polynomials for $a_1$:

\begin{equation}
C^{(0,m,m)}(a_1,a_2,a_3)  = \left({\cal N}^{(m)}(a_2){\cal N}^{(m)}(a_3)\right)^{1/2} \sum_\ell \frac{\chi_\ell(a_1){\cal U}_\ell^{(m)}(a_2,a_3)}{|\ell|_{q^r}}
\end{equation}
for some functions ${\cal U}_\ell^{(m)}(a_2,a_3)$. For future convenience, we have pulled out an $\ell-$independent normalization factor, with

\begin{equation}
{\cal N}^{(m)}(a)=1-\frac{a^2 q^{r-2m}+a^{-2} q^{2m}}{1+q^r}
\end{equation}
The functions ${\cal U}_\ell^{(m)}(a_2,a_3)$ possess some remarkable features. Computing them to several orders in a $q-$expansion we find that we can write them as

\begin{eqnarray}
{\cal U}_1^{(m)}(a,b)=1,~~~~~{\cal U}_\ell^{(m)}(a,b) = U_{\ell,1}^{(m)}(a)U_{\ell,1}^{(m)}(b)+U_{\ell,2}^{(m)}(a)U_{\ell,2}^{(m)}(b)
\end{eqnarray}
where $U_{\ell,1}^{(m)}(a)$ and $U_{\ell,2}^{(m)}(a)$ satisfy orthonormality properties with respect to the "measure" ${\cal N}^{(m)}(a)$:

\begin{eqnarray}
\label{ortho}
\frac{1}{2\pi i}\oint \frac{da}{ a}  {\cal N}^{(m)}(a) U_{\ell,i}^{(m)}(a) U_{\ell',j}^{(m)}(a) = \delta_{\ell \ell'} \delta_{ij}
\end{eqnarray}
Assuming that $U_{\ell,i}^{(m)}(a)$ contain only terms of the form $a^{-\ell},...,a^\ell$, any solution of (\ref{ortho}) leads to the correct values for ${\cal U}_\ell^{(m)}(a,b)$! However, such solutions are not unique. We found it convenient to choose the following basis
\begin{eqnarray}
U_{\ell,1}^{(m)}(a q^m)& =& \chi_\ell(a)\\
U_{\ell,2}^{(m)}(a q^m)& =& -i \frac{a^{1-\ell}((1-a^{2+2\ell})(q^r-q^{\ell r})+a^2(q^{(\ell+1)r}-1)+a^{2\ell}(1-q^{(\ell+1)r})) }{\sqrt{(1-q^{(\ell-1)r})(1-q^{(\ell+1)r})}(a^2-1)(a^2 q^r-1)}
\end{eqnarray}
from which the orthogonality conditions can be checked explicitly. The above definitions work for $0 < m <r/2$. For the special cases $m=0,r/2$ we obtain

\begin{eqnarray}
U_{\ell,1}^{(0)}(a)& =& U_{\ell,1}^{(r/2)}(a)=\chi_\ell(a)\\
U_{\ell,2}^{(0)}(a )& =& U_{\ell,2}^{(r/2)}(a )=0
\end{eqnarray}
The doubling of the base functions, from $\chi_\ell(a)$ to $U_{\ell,1}^{(0)}(a)$ plus $U_{\ell,2}^{(0)}(a)$, when $m \neq 0,r/2$, is expected, since for these cases, the gauge group is broken to the maximal torus.

Finally, let us mention that the orthonormality conditions (\ref{ortho}) imply
\begin{equation}
\frac{1}{2\pi i } \oint \frac{da'}{a'} {\cal N}^{(m)}(a') {\cal U}_\ell^{(m)}(a,a') {\cal U}_\ell^{(m)}(a',b) ={\cal U}_\ell^{(m)}(a,b)
\end{equation}

Let us now focus on the generic case in which $m_1,m_2,m_3 \neq 0$. For the case $r=1$, it is convenient to expand the structure constants in terms of Schur polynomials. In other words, given $C(a_1,a_2,a_3)_{r=1}$ we consider

\begin{equation}
C _{\ell_1,\ell_2,\ell_3}(a_1,a_2,a_3)_{r=1}=\int \prod_i [da_i] \chi_{\ell_i}(a_i) \chi_{\ell_i}(a_i')  C(a'_1,a'_2,a'_3)_{r=1}
\end{equation}
This action is diagonal in the index $\ell$ (see \cite{Gaiotto:2012xa} for a physical explanation of this fact), which leads to (\ref{qdef}) and manifests the structure behind S-duality. For the case $r>1$, it is then natural to follow the same procedure with the replacement

\begin{equation}
 \chi_{\ell}(a) \chi_{\ell}(a') \rightarrow   \left({\cal N}^{(m)}(a){\cal N}^{(m)}(a')\right)^{1/2}{\cal U}_\ell^{(m)}(a,a')
\end{equation}
Hence, given a general three-point function $C^{(m_1,m_2,m_3)}(a_1,a_2,a_3)$, we can consider the transformed one

\begin{equation}
C^{(m_1,m_2,m_3)}_{\ell_1,\ell_2,\ell_3}(a_1,a_2,a_3)= \int \prod_i [da'_i]  \left({\cal N}^{(m_i)}(a_i){\cal N}^{(m_i)}(a_i')\right)^{1/2}{\cal N}_{\ell_i}^{(m_i)}(a_i,a_i')  C^{(m_1,m_2,m_3)}(a'_1,a'_2,a'_3)
\end{equation}

Note that for the case $m=0$, we obtain the previously mentioned results, since $ \left({\cal N}^{(m)}(a){\cal N}^{(m)}(a')\right)^{1/2}{\cal U}_\ell^{(m)}(a,a')=  \chi_{\ell}(a) \chi_{\ell}(a') $. Quite surprisingly, we find that this action is again diagonal in the index $\ell$! Furthermore, the structure constant can be written as the sum of these components

\begin{eqnarray}
C^{(m_1,m_2,m_3)}(a_1,a_2,a_3) = \sum_\ell C^{(m_1,m_2,m_3)}_{\ell}(a_1,a_2,a_3) 
\end{eqnarray}
We define the three-point correlation functions of our putative 2d TFT as

\begin{eqnarray}
\langle V^{(m_1)}(a_1) V^{(m_2)}(a_2) V^{(m_3)}(a_3)  \rangle = \sum \langle V^{(m_1)}(a_1) V^{(m_2)}(a_2) V^{(m_3)}(a_3)  \rangle_\ell \\
\langle V^{(m_1)}(a_1) V^{(m_2)}(a_2) V^{(m_3)}(a_3)  \rangle_\ell  =\left(  \prod_{i=1}^3 {\cal N}^{(m_i)}(a_i)   \right)^{-1/2}   C^{(m_1,m_2,m_3)}_{\ell}(a_1,a_2,a_3) 
\end{eqnarray}
The normalization factor, is such that in the 2d TFT the gluing is done with the measure factor ${\cal N}^{(m)}(a)=1-\frac{a^2 q^{r-2m}+a^{-2} q^{2m}}{1+q^r}$. Note that the $m$ dependence of the measure factor can be absorbed by taking $a \rightarrow a q^m$. Furthermore, note that for $q \rightarrow 1$, the measure factor reduces to the usual $SU(2)$ Haar measure. 

Even though the correlation functions of the 2d TFT do not factorize into functions of the $a_i$, the orthonormality properties (\ref{ortho}) imply the following form

\begin{equation}
\label{2dpicture}
\langle V^{(m_1)}(a_1) V^{(m_2)}(a_2) V^{(m_3)}(a_3)  \rangle_\ell  = \sum_{i,j,k=1}^2 c^{({\cal R})}(q) f^{({\cal R})}_{\ell,ijk}(q) U_{\ell,i}^{(m_1)}(a_1) U_{\ell,j}^{(m_2)}(a_2) U_{\ell,k}^{(m_3)}(a_3)
\end{equation}
where we have stressed the fact that the functions $c^{({\cal R})}(q) $ and $f^{({\cal R})}_{\ell,ijk}(q)$ will depend on $q$. Besides the explicit dependence on the $m_i$ in the functions $U_{\ell,i}^{(m_i)}(a_i)$, the value of $f^{({\cal R})}_{\ell,ijk}(q)$ can change as we jump from one "region" to another. Here ${\cal R}$ runs over the possible regions and the values of the $m_i$ fix the region we are at. Let's for simplicity consider the case in which $r$ is odd and assume $0< m_1 \leq m_2 \leq m_3<r/2$. In this case:

\begin{equation}
{\cal R} =\left\{
\begin{array}{l l}      
     I & ~ m_1+m_2+m_2=r\\
   II & ~m_1+m_2=m_3 \\
   III & ~\mbox{other cases}
\end{array}\right.
\end{equation}
This division into regions is of course expected, due to the presence of $[{\bf m}.{\bf s}]_r$  in (\ref{Cvspoch}), namely, within a region we have a 'continuous dependence' on the $m_i$, but the expression 'jumps' when we cross to a different region. The normalization factor $c^{({\cal R})}$ has been pulled out in order to have $f^{({\cal R})}_{\ell=1,111}=1$ and is given by

\begin{equation}
c^{( I )}(q) = c^{( II )}(q) =\frac{1}{\sqrt{1+q^r}},~~~~~c^{( III )}(q) =\frac{1-q^r}{\sqrt{1+q^r}}
\end{equation}

The structure constants have certain symmetry properties under permutation of the holonomies $m_i$. This implies certain symmetries among the functions $f^{({\cal R})}_{\ell,ijk}$, namely $f^{(I)}_{\ell,ijk}$ and $f^{(III)}_{\ell,ijk}$ are invariant under permutation of $i,j,k$ and $f^{(II)}_{\ell,ijk}$ is invariant under the interchange of $i$ and $j$. Furthermore, up to a very high order in the $q-$expansion, we have checked the additional symmetries

\begin{eqnarray}
f^{(II)}_{\ell,111}= f^{(II)}_{\ell,122}=  f^{(II)}_{\ell,212}=f^{(III)}_{\ell,111}= f^{(III)}_{\ell,122}=  f^{(III)}_{\ell,212}= f^{(III)}_{\ell,221}\\
 f^{(II)}_{\ell,112}= f^{(II)}_{\ell,121}= f^{(II)}_{\ell,211}=  f^{(III)}_{\ell,112}= f^{(III)}_{\ell,121}= f^{(III)}_{\ell,211}=0
\end{eqnarray}
All in all, at each order $\ell$ and for generic $r$, the structure constants depend on eight functions of the fugacity $q$ (for small values of $r$ not all the regions are present). Let us introduce the following notation

\begin{eqnarray}
h_\ell^{(1)}(q)& =& f^{(I)}_{\ell,111}\\
h_\ell^{(2)}(q) &=& f^{(I)}_{\ell,112} \mbox{~~(plus permutations)}\\
h_\ell^{(3)}(q) &=& f^{(I)}_{\ell,122} \mbox{~~(plus permutations)}\\
h_\ell^{(4)}(q) &=& f^{(I)}_{\ell,222}\\
h_\ell^{(5)}(q) &=& f^{(II)}_{\ell,111}= f^{(II)}_{\ell,122}=f^{(II)}_{\ell,212} =f^{(III)}_{\ell,111}= f^{(III)}_{\ell,122}   \mbox{~~(plus permutations)}\\
h_\ell^{(6)}(q) &=&   f^{(II)}_{\ell,221}\\
h_\ell^{(7)}(q) &=&  f^{(II)}_{\ell,222}\\
h_\ell^{(8)}(q) &=&  f^{(III)}_{\ell,222}
\end{eqnarray}
Up to very high powers of the fugacity, and for several values of $\ell$ and $r$, we have found the following expressions for these functions 
\begin{eqnarray}
h_\ell^{(1)}(q)& =&  q^{-1/2(\ell-1)r} \frac{(1-q^r)(1+q^{\ell r}+q^{2 \ell r}-q^{(1+\ell)r})}{(1+q^r)(1-q^{\ell r})}\\
h_\ell^{(2)}(q) &=& i q^{-1/2(\ell-1)r}  \frac{(1-q^r)(1+q^{\ell r})}{(1+q^r)(1-q^{\ell r})} \sqrt{1-q^{(\ell-1)r}+q^{2\ell r}-q^{(\ell+1)r}}\\
h_\ell^{(3)}(q) &=& -q^{-1/2(\ell-1)r} \frac{(1-q^r)(1-q^{(\ell-1) r}+q^{\ell r}+q^{ 2\ell r})}{(1+q^r)(1-q^{\ell r})}\\
h_\ell^{(4)}(q) &=& - i q^{-1/2(\ell-1)r} \frac{(1-q^r)(1+q^{\ell r})}{(1+q^r)(1-q^{\ell r})} \frac{1-2 q^{(\ell-1)r}+q^{2\ell r}}{\sqrt{1-q^{(\ell-1)r}+q^{2\ell r}-q^{(\ell+1)r}}}\\
h_\ell^{(5)}(q) &=& q^{1/2(\ell-1)r}\frac{1-q^r}{1-q^{\ell r}}  \\
h_\ell^{(6)}(q) &=& -q^{1/2(\ell+1)r}\frac{1-q^r}{1-q^{\ell r}} \\
h_\ell^{(7)}(q) &=&i q^{1/2(\ell-1)r} \frac{(1-q^r)^2(1+q^{\ell r})}{(1-q^{\ell r})} \frac{1}{\sqrt{1-q^{(\ell-1)r}+q^{2\ell r}-q^{(\ell+1)r}}}\\
h_\ell^{(8)}(q) &=& 2i q^{1/2(\ell-1)r} \frac{(1-q^r)(1+q^{\ell r})}{(1-q^{\ell r}) \sqrt{(1-q^{(\ell-1)r})(1-q^{(\ell+1)r})}}
\end{eqnarray}
These expressions are also valid for $r$ even, except the case where $m_1+m_2=m_3=r/2$ (which can only exist for $r$ even). If we call that region ${\cal R}=IV$, we find 

\begin{eqnarray}
f^{(IV)}_{\ell,111} &=& - f^{(IV)}_{\ell,221} = \frac{(1-q^r)(1+q^{\ell r})}{(1+q^r)(1-q^{\ell r})} \\
f^{(IV)}_{\ell,121} &=& f^{(IV)}_{\ell,211} = i \frac{(1-q^r)\sqrt{1-q^{(\ell-1)r} -q^{(\ell+1)r}+q^{2\ell r}}}{(1+q^r)(1-q^{\ell r})} 
\end{eqnarray}
as well as $c^{(IV)}(q)=1$. Remember that if $m_3=r/2$, the third index of $f_{\ell,ijk}$ is forced to be one. Hence, we have fully defined the structure constants of the putative 2d TFT. Equivalently, we can say that these are the partition functions of the 2d TFT on a sphere with three punctures, where the punctures have labels $(a_i,m_i)$.

\subsection{Crossing symmetry}
As mentioned above, the structure constants correspond to the partition function of a 2d TFT on a sphere with three punctures. The partition function on a generic Riemann surface  with punctures can be computed by the gluing procedure, decomposing the Riemann surface into pairs of pants joined by tubes. A novel feature, not present in usual 2d TFT, is the sum over the holonomy $m$ of the intermediate state. The factorization (\ref{2dpicture}) implies a result of the form

\begin{equation}
Z(a_1,m_1,...,a_n,m_n) = \sum_\ell \sum_{i_1,...,i_n} f_{\ell,i_1,...,i_n}^{m_1,...,m_n}(q)U_{\ell,i_1}^{(m_1)}(a_1)...U_{\ell,i_n}^{(m_n)}(a_n)
\end{equation}
where the $q$-dependent factors  $f_{\ell,i_1,...,i_n}^{m_1,...,m_n}(q)$ can be easily computed from the formulae given in this paper. For the gluing procedure to be consistent, it should not matter which particular pants decomposition we choose in order to do the computation. From the point of view of the 4d index, this is true once we assume S-duality. From the 2d TFT perspective, this happens provided the four-point correlation function satisfies crossing symmetry, see figure 1.

\begin{figure}[h]
\centering
\includegraphics[scale=0.5]{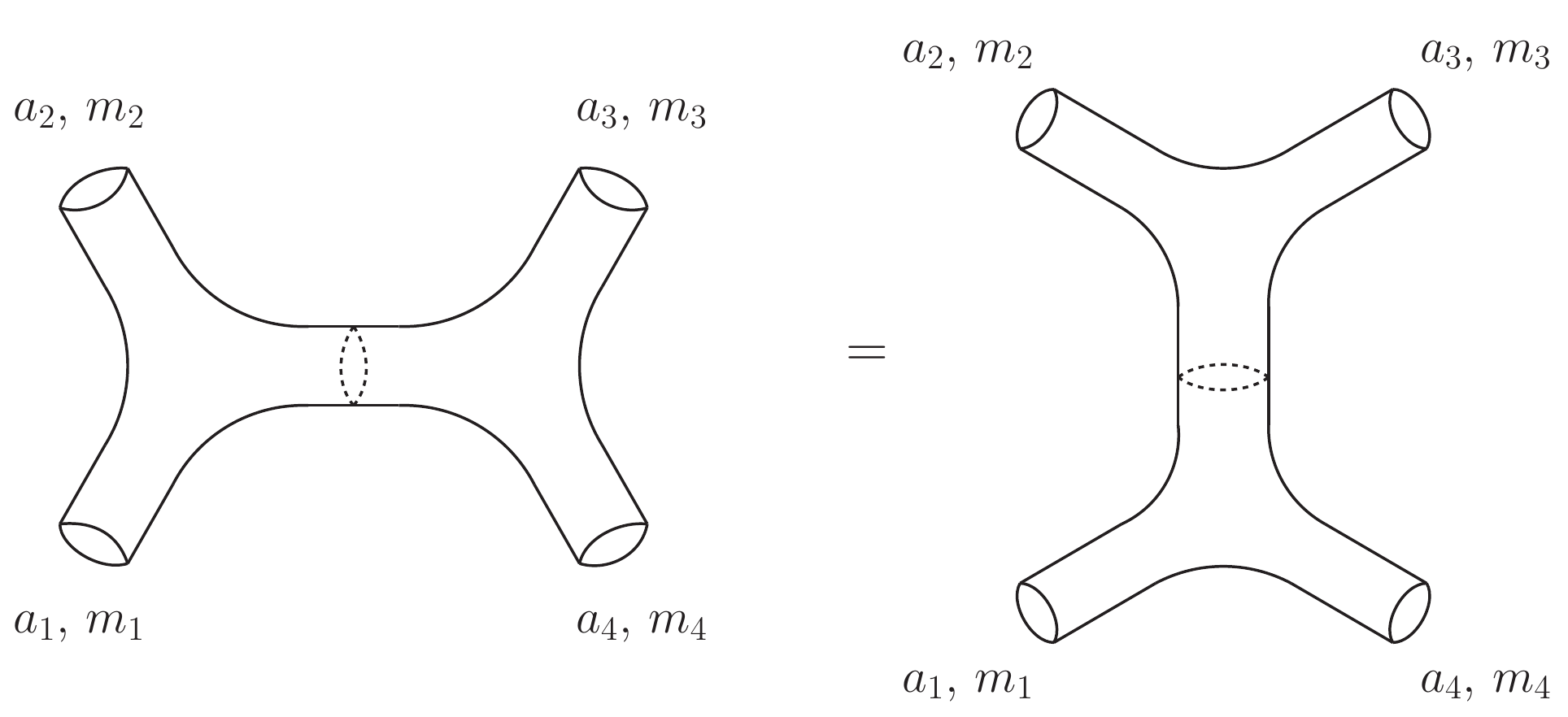}
\caption{The four point correlation function should be crossing symmetric. On both sides one should sum over the holonomy $m$ of the internal state. }
\end{figure}

 More precisely, we should have

\begin{eqnarray}
\sum_m \int [da]^m \langle V^{(m_1)}(a_1) V^{(m_2)}(a_2) V^{(m)}(a)  \rangle  \langle V^{(m)}(a) V^{(m_3)}(a_3) V^{(m_4)}(a_4)  \rangle =\\
\sum_m \int [da]^m \langle V^{(m_1)}(a_1) V^{(m_3)}(a_3) V^{(m)}(a)  \rangle  \langle V^{(m)}(a) V^{(m_2)}(a_2) V^{(m_4)}(a_4)  \rangle
\end{eqnarray}
where the sums over $m$ run over the holonomies compatible with the selection rules of the putative 2d TFT, discussed above, and the integration over $a$ is done with the measure ${\cal N}^{(m)}(a)$. The factorization (\ref{2dpicture}) implies that crossing symmetry should be true at each level $\ell$ independently. For a fixed choice of the external holonomies $m_{1,2,3,4}$, crossing symmetry holds provided the functions $h_\ell^{(i)}(q)$ satisfy specific quadratic constraints. When $r$ is large enough, the number of constraints is much larger than the number of functions . It turns out that one can isolate seven "basic" constraints: 

\begin{eqnarray*}
h_\ell^{(1)}(q) + h_\ell^{(3)}(q) = \frac{1-q^r}{|\ell|_{q^r}},~~~~~h_\ell^{(2)}(q) ^2= \frac{(1- |\ell|_{q^r} h_\ell^{(3)}(q) )(q^r+ |\ell|_{q^r} h_\ell^{(3)}(q) )}{|\ell|_{q^r}^2} \\
h_\ell^{(4)}(q) h_\ell^{(2)}(q)=h_\ell^{(3)}(q)^2-\frac{1}{|\ell|_{q^r}^2} ,~~~~~  h_\ell^{(5)}(q) |\ell|_{q^r} =1,~~~~~h_\ell^{(6)}(q) +\frac{q^r}{ |\ell|_{q^r} }=0\\
h_\ell^{(7)}(q)^2 (q^r+ h_\ell^{(3)}(q) |\ell|_{q^r}) |\ell|_{q^r}^2=(1-q^r)^2(1- h_\ell^{(3)}(q) |\ell|_{q^r}),~~~~~ h_\ell^{(7)}(q)=\frac{1}{2}(1-q^r) h_\ell^{(8)}(q)
\end{eqnarray*}
which imply all the constraints for any values of $0 \leq m_{1,2,3,4} < r/2$. One can explicitly check that the functions $h_\ell^{(i)}(q)$ found in this paper, indeed satisfy these constraints. As a result, correlation functions of this 2d TFT possess crossing symmetry and the super-conformal index of super-symmetric gauge theories on Lens spaces, in the fugacity slice considered in this paper, possesses S-duality.

\section{Conclusions}

In this paper we have considered the super-conformal index of four dimensional ${\cal N}=2$ gauge theories on $S^1 \times L(r,1)$, where $L(r,1)$ is a Lens space. The index depends on three fugacities $(p,q,t)$ and we have focused on a one-parameter family of the form $(p,q,t)=(0,q,q^r)$. For $r=1$ this reduces to the so called Schur limit, studied in \cite{Gadde:2011ik}. In addition to the fugacities, the index also depends on $SU(2)$ holonomies $a_i$ and holonomies $m_i$ along $S^1_H$. One interesting feature of the $p \rightarrow 0$ limit, is that structure constants satisfy a selection rule on the $m_i$'s, analogous to the selection rules for $SU(2)_r$. This feature however, will not persist in general. In this limit we managed to write the structure constants in a way that resembles the partition function of a 2d TFT on a sphere with three punctures, see (\ref{2dpicture}). Such a factorization is written in  terms of a basis of functions, and $q$-dependent coefficients $h_\ell^{(i)}$. The coefficients computed in this paper satisfy a basic set of constraints which imply crossing symmetry for higher point correlators. Crossing symmetry translates into S-duality for the super-conformal index under consideration.  

There are several open problems. First, it would be interesting to identify the relevant 2d TFT and understand the precise meaning of the holonomy $m$ in the 2d picture. One way to proceed would be along the lines of \cite{Fukuda:2012jr}, by considering the compactification of 5d YM on $S^3/\mathbb{Z}_r$. Alternatively, one could try to understand the relation to the correspondence between 4d ${\cal N}=2$ partition functions on spaces of the form $\mathbb{R}^4/\mathbb{Z}_p$ and 2d Para-Liouville/Toda theories, see {\it e.g.} \cite{Belavin:2011pp,Nishioka:2011jk,Bonelli:2011jx}. A simpler problem would be to try to identify the 2d theory in the ``undeformed" case $q \rightarrow 1$. In this case, several of our expressions simplify considerably. For $r=1$ the relevant 2d theory is $q-$deformed Yang-Mills in the zero area limit, which reduces to usual 2d YM in the limit $q=1$.

Another interesting problem would be to give a physical interpretation for the functions ${\cal U}_\ell(a,b)$ found in this paper (or their components). For instance, for the $r=1$ case it was observed \cite{Gaiotto:2012xa} that the structure constants factorize into eigenfunctions of certain polynomials. It would be interesting to extend their analysis to the present case. 

Finally, our results can be extended in several directions. One can try to include the other fugacities, for instance by focusing on the ``Macdonald" limit of \cite{Gadde:2011uv}. As already mentioned, in this limit we will still have a selection rule for the $m_i$'s. It would be interesting to understand precisely the relation between enhanced super-symmetry and this selection rule. Finally, we could try to extend our results to Lens spaces of the form $L(p,q)$. For this one would first need to compute the super-conformal index in such spaces. As explained in \cite{Alday:2012au}, the treatment in this case is more involved.

\section*{Acknowledgments}

We would like to thank Y. Tachikawa for interesting discussions and M. Yamazaki and F. Benini for detailed explanations of \cite{Benini:2011nc}.
The work of L.F.A. and M.F. is supported by ERC STG grant 306260. The work of M.B. is supported by the EPSRC.

  \end{document}